\documentstyle[epsf,psfig,natbib_mod]{mn2e}

\newcommand{\ltsimeq}{\raisebox{-0.6ex}{$\,\stackrel 
        {\raisebox{-.2ex}{$\textstyle <$}}{\sim}\,$}} 
\newcommand{\gtsimeq}{\raisebox{-0.6ex}{$\,\stackrel 
        {\raisebox{-.2ex}{$\textstyle >$}}{\sim}\,$}} 

\begin{document}
\title[Probing dark energy with baryonic oscillations and future radio 
surveys]{Probing dark energy with baryonic oscillations and 
future radio surveys of neutral Hydrogen.}
\author[Abdalla \& Rawlings]
{F. B. Abdalla$^{1}$,
S. Rawlings$^{1}$.
\\
$^{1}$Department of Physics, Oxford University,
Denys Wilkinson Building, Keble Road, Oxford OX1 3RH, U.K.\\
}
\maketitle

\begin{abstract}
Current surveys may be on the verge of 
measuring the baryonic oscillations in the galaxy power spectrum which are clearly seen
imprinted on the Cosmic Microwave Background. It has recently been proposed that
these oscillations allow a `standard ruler' method of probing the equation of state 
of dark energy. In this paper we present a new calculation of the 
number of galaxies future radio telescopes will detect in surveys of the 
sky in neutral Hydrogen (HI). 
We estimate the likely statistical errors if the standard 
ruler method were to be applied to such surveys. We emphasise uncertainties in our 
calculations, and pinpoint the most important features of future HI surveys
if they are to provide new constraints on dark energy via baryonic oscillations.
Designs of future radio telescopes are required 
to have a large bandwidth (characterised by $\beta$, 
the ratio of the instantaneous bandwidth to the bandwidth required by survey)
and to have the widest instantaneous (1.4~GHz) field of view ($FOV$) possible.
Given the expected sensitivity of a future
Square Kilometre Array (SKA), given that half of its collecting area
will be concentrated in a core of diameter $\sim 5 ~ \rm km$, and 
given a reasonable survey duration ($T_0$ $\sim$ 1 yr), 
we show that there will be negligible shot noise on a power spectrum derived from
HI galaxies out to redshift $z \simeq 1.5$. To 
access the largest cosmic volume possible by surveying 
all the sky available, we
argue that $\beta$, $T_0$ 
and $FOV$ must obey the relation 
$\beta \, FOV \, T_0 \gtsimeq 10 \, \rm deg^2$ yr.  
An $\sim$1-yr 
SKA survey would then contain
$\gtsimeq 10^{9} (f_{\rm sky}/0.5)$ HI galaxies and 
provide constraints on the dark-energy parameter $w$ of order 
$\Delta w \simeq 0.01 (f_{\rm sky}/0.5)^{-0.5}$, 
where $f_{\rm sky}$ is the fraction of the whole sky observed.
\end{abstract}


\section{Introduction}

We are now widely believed to have entered an era of precision 
cosmology (e.g. Percival et al. 2001; Spergel et al. 2003). It is 
therefore important that all new surveys, and all new equipment designed to 
make these surveys, are able to make precision measurements. 
These measurements should not only improve on the current constraints on 
the cosmological parameters but also begin to seriously constrain 
the equation of state of dark energy and its evolution with cosmic epoch.
These constraints will eventually distinguish between a cosmological constant and other
models for dark energy such as quintessence 
(e.g. Carroll, Press \& Turner 1992; Caldwell, Dave \& Steinhardt 1998). 

Recently, much effort has been expended to establish the best 
way of determining the properties of dark energy. Several methods have 
been proposed: the use of Type Ia supernovae to 
probe the luminosity distance (e.g. Weller \& Albrecht 2002); the 
use of cluster number counts (Haiman, Mohr \& Holder 2001) or counts of 
galaxies (Newman \& Davis 2000); weak gravitational lensing 
(Cooray \& Huterer 1999); the Alcock-Paczynski test 
(Ballinger, Peacock \& Heavens 1996); 
and the CMB (e.g. Douspis et al. 2003). In this paper we examine one method in particular, 
the `standard ruler' method based on baryonic 
oscillations (Eisenstein 2002), as several authors 
(Blake \& Glazebrook 2003; Hu \& Haiman 2003; Seo \& Eisenstein 2003)
have argued that it suffers from a set of systematic errors that are 
much less serious than those of the other methods.

We investigate here the r\^{o}le in dark energy studies
of future radio surveys of neutral Hydrogen (HI).
Such surveys are likely to reach full fruition with the proposed next-generation
radio synthesis array, the Square Kilometre Array 
(SKA; Carilli \& Rawlings 2004). 

In Sec.2 we describe how we would be able to detect HI
at high redshifts with future radio surveys. There are no direct 
observations of HI in emission in the high-redshift Universe as the current radio telescopes 
used to search for HI in emission are only sensitive enough to reach 
redshifts of around 0.2 (Zwaan, van Dokkum \& Verheijen 2001). 
Nevertheless, we have evidence of large amounts of high-redshift HI through the 
damped-Ly$\alpha$ objects seen in absorption in quasar optical spectra 
(Storrie-Lombardi \& Wolfe 2000; Peroux et al. 2001). 
In Sec.3 we use this information to constrain
possible evolutions of the HI mass function and produce a new calculation of
the number density of HI galaxies to be 
detected by future radio telescopes which improve on
previous estimates (Briggs 1999; van der Hulst 1999).
We compare our `best guess'
evolutionary model with other observational constraints in Sec.4; fitting formulae are 
given in Appendix A1.

Having an estimate of what future radio surveys will be able to see 
in the 21-cm line of HI in emission, we can see what cosmological 
tests we can perform on such data and decide on their pros and cons.
We focus here on probing dark energy with the baryonic oscillations
method but other cosmological experiments
are possible (see Blake et al. 2004; Rawlings et al. 2004).   
In Sec.5 we show that, given the likely capabilities of future radio telescopes, 
the optimal survey would be an `all hemisphere survey' of all the sky
area available.
We then compute what comoving cosmological volume and numbers of sources 
are likely to be available 
in such surveys and we estimate the accuracy that the baryonic oscillations
`standard ruler' 
method (e.g. Blake \& Glazebrook 2003; Hu \& Haiman 2003; Seo \& Eisenstein 2003) 
can give us in measuring the equation of state of dark energy; 
this is typically described by the parameter $w=p/ \rho$ (Turner \& White 1997), 
where $p$ is the pressure and $\rho {\rm c}^2$ is the energy 
density of the dark energy component.

In Sec.6 we discuss the uncertainties of our approach and discuss how they might influence 
the results of the standard ruler method used. We also discuss how the results 
would change if different assumptions are made for key features of the future radio 
surveys as well as some potential 
problems in using this method with future radio survey data.

For this paper we adopt the following cosmological values: $\Omega_M=0.3$, 
$\Omega_\Lambda=0.7$ and ${\rm h_{70}}=1$. 
We use the matter power spectrum given in Bardeen et al. (1986) with a normalisation given
by the WMAP results (Spergel et al. 2003) which corresponds to $\sigma_8\simeq0.84$.
For a given type of matter $x$ we define $\Omega_x$ as being the ratio of the density of $x$ 
to the critical density of the Universe today. 
When we refer to volumes, lengths, etc, we consider comoving
cosmological values unless specified otherwise. 

Unless stated otherwise if we mention the
field of view ({\it FOV}) of a radio telescope/array we are referring to the instantaneous 
field of view this instrument possesses, and can image, at 1.4 GHz.
It is vital to remember that for many radio telescopes the field of 
view which can be imaged gets larger at lower frequencies.

\section{Prospects for future radio survey of HI}

\subsection{Sensitivity of radio receivers}

The ratio of the signal to the noise power in a single-polarisation radio receiver is

\begin{equation}
\frac{\frac{1}{2}A_{eff} S \Delta \nu}{k T_{sys} \Delta \nu}=\frac{A_{eff}S}{2 k T_{sys}}
\end{equation}

\noindent where $A_{eff}$ is the effective collective area of the telescope 
(incorporating all inefficiencies), $S$ is the flux density, $\Delta \nu$ is 
the bandwidth and $T_{sys}$ is the system temperature (incorporating all contributions).

In this paper we will scale all limiting sensitivities to that expected for the SKA.
The SKA science requirements (Jones 2004) demand
$A_{eff} / T_{sys} = 2 . 10^4 {\rm m}^2 / {\rm K} $ over the frequency range
0.5 to 5 GHz. As the discussions in this paper will be limited to HI at redshifts 
$z \ltsimeq 2$, i.e. frequencies in the range 0.5 to 1.4 GHz, this means that we can 
write the `radiometry equation' for the SKA in a very simple form,
namely

\begin{equation}
S_{lim}=\frac{2 k T_{sys}}{A_{eff}\sqrt{2\Delta \nu t}} \simeq \frac{100 \, {\rm nJy}}{\sqrt{\Delta \nu t}}
\end{equation}

\noindent where $S_{lim}$ is the rms sensitivity for dual-polarisation observations with the SKA and the 
$\sqrt{2\Delta \nu t}$ term allows for the increase in sensitivity by averaging
independent measurements of the signal-to-noise ratio.

\subsection{Mass detection limit of HI}
Neutral Hydrogen (HI) will be found in emission with future radio surveys via the
21-cm line radiation due to the difference in energy in hyperfine atomic
structure (e.g. Field 1958). From atomic physics we know that the emissivity is

\begin{equation}
\epsilon_{\nu}=\frac{1}{4\pi}h\nu_{12} {\rm A}_{12}
\frac{N_2}{N_{\rm H}}N_{\rm H} \varphi(\nu) ,
\end{equation}

\noindent where ${\nu}_{12}$ and ${\rm A}_{12}$ are 
the rest-frame frequency and  
Einstein A coefficient for this transition respectively, $\varphi(\nu)$ 
is the line profile of the 21-cm line which is considered here as a delta
function, and $N_{\rm H}$ and $N_2$ are the total number of Hydrogen atoms, 
and number of atoms in the upper (level 2) respectively. 

We can write $N_2/N_{\rm H} =({N_2}/{N_1})/({N_1}/{N_{\rm H}})$. The first ratio is given by 
${N_2}/{N_1}=({g_2}/{g_1})exp({-{h\nu_{12}}/{T_{\rm S}}})$, where $N_1$ is the number of 
atoms in level 1 and $g_1=1$ and $g_2=3$ are the statistical weights for these levels. 
$T_{\rm S}$ is the so called spin 
temperature and is an effective temperature resulting from the coupling of the CMB
temperature $T_{\rm CMB}$ and the kinetic temperature $T_{\rm K}$ of the gas. 
In the case of the CMB radiation alone the spin temperature will equal the CMB temperature.
If we have any collisional excitation or scattering by Ly$\alpha$ photons 
the spin temperature will couple to the kinetic temperature as well as the CMB temperature
and will therefore be a weighted average of both 
(Rohlfs \& Wilson 1999). In cases we are considering here (dense clouds) we have both the 
kinetic temperature and the Cosmic Microwave Background (CMB) 
temperature much larger than $h\nu_{12}=0.06 \, {\rm K}$ 
and, given that ${g_2}/{g_1}=3$, we have ${N_2}/{N_1}\simeq 3$. 

Observations show that $T_{\rm S}$
can be as large as 300 K in low-redshift galaxies (Chengalur \& Kanekar 2000) and 
that in damped-Ly$\alpha$ objects at higher redshifts, limits 
on HI absorption lines imply larger values of $T_{\rm S}$ 
of order of 1000 K or more (Kanekar \& Chengalur 2003). 
If we are dealing with HI in emission 
we will get the same signal whatever $T_{\rm S}$, however the fact 
that the spin temperature is higher at higher redshifts is telling us that we are 
probably probing a different kind of interstellar medium. At high redshift we are
typically probing 
a warm neutral intergalactic medium that is present in larger fraction in 
smaller less dense dark-matter halos (Young \& Knezek 1989). 
We can safely say that in all cases of our interest ${N_2}/{N_{\rm H}}\simeq{3}/{4}$.

If we consider a cloud of Hydrogen, the monochromatic luminosity we would get from 
this 21-cm line emission will be

\begin{equation}
L_\nu= \int_V\int_\Omega\epsilon_{\nu} {\rm d}V {\rm d}\Omega
=\frac{3}{4}h\nu A_{12}\frac{M_{\rm HI}}{m_{\rm H}}\varphi(\nu) .
\end{equation}

So, given the expression for the monochromatic flux density 
$S_{\nu}={L_{\nu (1+z)}(1+z)}/{(4\pi D_L^2(z))}$
(Peacock 1999; Eqn 3.87), 
where $D_L(z)$ is the luminosity distance to the galaxy,
we can integrate Eqn.4 over frequency to obtain

\begin{equation}
\int S_{\nu} {\rm d}\nu= \frac{1}{4\pi} \frac{3}{4} hA_{12} \frac{M_{\rm HI}}{m_{\rm H}}
\frac{1+z}{D_L^2(z)} \int \nu \, \varphi(\nu) \, {\rm d}\nu ,
\end{equation}

\noindent and therefore obtain the expression for the mass corresponding to the flux 
seen in our observations, namely

\begin{equation}
M_{HI}(z)=\frac{16\pi}{3}\frac{m_{\rm H}}{A_{12}hc}\frac{D_L^2(z)}{1+z}\int S_{\nu} \, {\rm d}V,
\end{equation}

\noindent where the integral is now over $V$, the line-of-sight width
corresponding to the projected circular velocity of the galaxy. In more 
useful units

\begin{equation}
\frac{M_{HI}(z)}{M_\odot}=\frac{0.235}{1+z} \frac{D_L^2(z)}{\rm Mpc^{2}} \frac{S_{\nu}}{\rm \nu Jy} \frac{V}{{\rm km/s}}
\end{equation}

\noindent noting that the factor of (1+z) arises as $S_{\nu} V$ needs to
be multiplied by $\nu_{12}/(1+z)$ to produce an integrated line flux.

We have neglected HI self-absorption effects which means that the HI mass may be 
a slight underestimate but this is likely to be a problem 
only when the disks of the largest galaxies seen close to edge on 
(Rao, Turnshek \& Briggs 1995).

\subsection{Sensitivity limits of future radio surveys}

The HIPASS survey (Ryan-Weber et al. 2002), used channels of velocity width
$\Delta V = 13$ km/s and was capable of detecting typical galaxies out to $z \simeq 0.02$. 
It is important that the $\Delta V$ chosen 
for an HI survey is not larger than the 
velocity width of the object being observed because this would 
result both in loss of signal-to-noise ratio and the danger of mistaking 
signal for interference.
The HIPASS survey does not show any evidence that 
many sources have low velocity width and the very lowest velocities found are 
around 30 km/s. These widths might change systematically with redshift. 
Zwaan et al. (2001) have
detected an example of a HI rich cloud at $z=0.18$ with a velocity profile
of width $V=$ 60 km/s.
We take $\Delta V =$ 30 km/s but
caution that even finer velocity bins may prove necessary to avoid losing
signal-to-noise ratio on the narrowest-line objects, particularly if line width
correlates negatively with redshift. The current 
`Strawman Design' for the SKA (Jones 2004) suggests
channels of width $\Delta V \simeq \, 30 \,{\rm km/s}$ will be available. 
We assume throughout that HI lines are detected and measured using 
optimal smoothing techniques.

In the standard picture of galaxy formation 
(e.g. Rees \& Ostriker 1977; White \& Rees 1978), 
we expect dark-matter halos to form potential wells, with 
gas falling into these potential wells becoming shock heated to 
the virial temperature. Cooling can then occur if the free-fall 
timescale is longer than the cooling timescale. 
Dekel \& Silk (1986) have showed that the objects that can cool have 
circular velocities in between 10 km/s and 200 km/s. Further to this,
it is argued that objects with circular velocity between 10 km/s and 30 km/s
are likely to be 
totally dark as their potential wells are so shallow that the cold gas will
disappear by evaporation due to photo-ionisation (Dekel 2004). 
These theoretical arguments lead us to do 
our calculations with an assumed $\Delta V$ = 30 km/s
which means that only objects with the lowest circular 
velocities and with nearly face-on rotating disks
will have HI 
line profiles with $V \ltsimeq \Delta V$.

In every $\Delta V$ = 30 km/s 
channel there will be a root-mean-square noise that will be dependent on frequency $\nu$, 
estimated for the SKA to be  $\sim$ 2 $\mu$Jy at $\nu=1.4$ GHz (HI at $z=0$) to 
$\sim$ 4 $\mu$Jy at $\nu=470$ MHz (HI at $z \simeq 2$) for a 4-hour 
pointed observation. We denote this noise by $\sigma_{\rm 4h}$.
 
We assume that the average Hydrogen-rich galaxy has a rectangular 
line-of-sight velocity spread 
$V_0=300$ km/s (which corresponds to a circular velocity of around 200 km/s) 
at $z=0$ and we also assume a 
`Tully-Fisher-like' relationship $V^4 \propto M_{\rm DM}^2 / R_0^2$ 
(Peacock 1999; p.622) that would hold at higher redshift, where $M_{\rm DM}$ 
is the dark-matter mass of the galaxy and $R_0$ is the galaxy radius. 
The evolution we choose for $M_{\rm DM}$ and $R_0$ with redshift is explained in Sec. 3.2
and this will impose a corresponding scaling of $V$ with $z$; this 
choice will in fact force the line width to correlate negatively with redshift 
(as $V(z)=V_{0} ({1+z})^{-1/2}$) although the physical lower limit of 30 km/s
proposed by Dekel (2004) should mean that this cannot decrease 
without limit for halos containing HI. 
Then the limiting HI mass that a radio survey will be able to detect at 
redshift $z$ is

\begin{equation}
M_{\rm HI}(z)=\frac{16\pi}{3}\frac{m_{\rm H}}{A_{12}hc}\frac{D_L^2(z)}{1+z} {f}^{-1}
\frac{V(z)}{\sqrt{V(z)/{\Delta}V}} S_{\rm N} \sigma_{\rm 4h}
\sqrt{\frac{4}{t}},
\end{equation}

\noindent where 
$S_{\rm N}$ is the signal-to-noise 
level we choose to yield a robust detection, $t$ is 
the integration time in hours for a given $FOV$ 
and $f$ is the fraction of the sensitivity
relative to the SKA; by definition $f=1$ for the SKA, 
and current radio telescopes have $f \ltsimeq 0.01$. 

\begin{figure}
\begin{center}
\centerline{\psfig{file=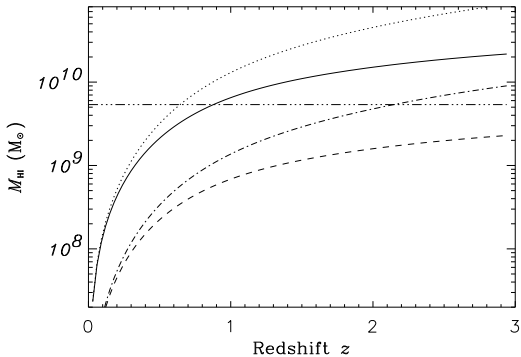,width=7.5cm,angle=0,clip=}}
\caption { Limiting HI mass $M_{\rm HI}$ for 
surveys with an SKA-like instrument for a signal-to-noise ($S_{\rm N}$) ratio of 10 in a 
4-hour integration time (solid line and dotted line) and a 360-hour
integration time (dot-dashed line and dashed line); the mass is in units of $M_\odot$.
The dotted and dot-dashed lines assume pointed observations and the 
solid and dashed lines assume tiled surveys as discussed in Sec 2.4.
The horizontal line corresponds to the break of the HI mass 
function $M_{\rm HI}^{\star}$ at low redshifts from Zwaan et al. (2003). 
We can see how an SKA 360-hour integration time can
take us very deep in $z$, but a simple 4-hour SKA integration time is enough to detect
an $M_{\rm HI}^{\star}$ galaxy (assuming no evolution in the break of the HI mass function)
out to redshift $z \sim 1$. This is for an array that has an effective field of view scaling 
with frequency $\nu$ as $\nu^{-2}$ (see Secs. 2.2 and 2.3).}
\label{mass_limit}
\end{center}
\end{figure}

\subsection{Survey geometry}

Currently the `Strawman Design' for the SKA (Jones 2004) 
has a {\it FOV} of 
at least 1 ${\rm deg}^2$. For most realisations of an SKA, 
the field of view will be much larger at frequencies smaller than 1.4 GHz, which will 
correspond to HI at redshifts larger than $z\simeq 0$. 
In fact the field of view, in units of ${\rm deg}^2$, 
will typically (e.g. because it is controlled by the diffraction limit of a dish) 
grow as $(1+z)^2$ if we are probing HI at increasing redshift.
In Sec.5 we will consider future telescopes where the field of view for
HI galaxies may vary with a different power of the frequency/redshift.

So, let us consider for illustrative purposes that we have a square beam and that we would 
like to cover a square sky patch of 64 ${\rm deg}^2$ with an integration time of 
8 hours per square degree. If we simply point the telescope 64 times at each square that we 
will name A1,A2...H7,H8 then we will have covered the sky smoothly at $z=0$ but
the data at $z=0.5$ will have parts of the sky that will have a higher sensitivity 
than others. This would be an undesired feature in the data for the purposes of making 
a uniform survey of HI.

In order to deal with this, we consider the following. 
Instead of pointing the telescope at each of A1,A2,... we take data {\it n} times 
in between A1 and A2 with 1/{\it n} th of the total time we would have spent 
on each of A1 and A2. We can then take the data that we receive 
from each small pointing and add it to the data available 
from other pointings. We end up with a survey 
with increasing sensitivity for increasing redshifts because a source at higher
redshift will be accessible to a larger fraction of the pointings and will therefore 
have a longer effective integration time for higher-redshift objects. In fact the effective 
integration time for such a `tiled' survey will increase smoothly as 
$(1+z)^2$ for a given integration time at $z=0$.
Ideally we would like to have a very smooth survey but in practise it may not 
possible to obtain maps with very large {\it n} because of limited computing capabilities. 
The wiggles on the power spectrum are at intervals of $\sim 0.05$ ${\rm Mpc}^{-1}$ 
so the survey needs to have a smooth window function even on scales corresponding to
$k \sim 0.01 \, {\rm Mpc}^{-1}$. If this is not the case the wiggles will be 
smoothed out by correlated errors on the power spectrum estimation
(see Blake \& Glazebrook 2003). Thus for the purposes of this experiment
we would like to have a smooth sky map on sizes of $\gtsimeq 600$ Mpc. 
The choice of $n \gtsimeq 10$ would ensure that smoothness is achieved on scales 
of the same size as the wiggles at the redshifts ($z \sim 1$) of interest.
We also would not have excessive data storage requirements as the integration time would 
be of the order of minutes similar to those used in current radio surveys.
More complicated survey schemes will be needed
with interferometric arrays to ensure optimal UV coverage.

So, for a given time of survey per square degree the limiting mass that the SKA 
will be able to see is 

\begin{equation}
M_{\rm HI}(z)=\frac{16\pi}{3}\frac{m_{\rm H}}{{\rm A}_{12}hc}\frac{D_L^2(z)}{(1+z)^{1+p}} f^{-1}
\sqrt{V(z) {\Delta}V} S_{\rm N} \sigma_{\rm 4h} 
\sqrt{\frac{4}{t}},
\end{equation}

\noindent where we have assumed that all redshifts are accessible by a single pointing and
where $p$ is defined by the field of view changing with frequency $\nu$
as $\nu^{-2p}$; i.e. for a $\nu^{-2}$ dependence we have $p=1$.

As we are covering a certain patch of the sky, 
the centre of this patch will have the sensitivity given by Eqn. 9, but 
the corners of the survey 
will have lower sensitivities as the beam will not have covered those areas as often 
as those in the centre. This will be the case for most of the survey 
area if we are looking at a small patch of the sky but if we are performing a large survey 
then this area with smaller sensitivity will be a small fraction of the total area of 
the survey, and can be neglected. We will show in Sec.5 that the optimal survey for this
experiment is an all-sky-survey, so this effect should therefore be negligible.

The assumption that the field of view changes with redshift will make a big
difference to the total 
cosmic volume being surveyed in a given length of time and this assumption is
relaxed in Sec.5 where we will consider different values of $p$.

\subsection{Source count}

We consider the local mass function of HI (Zwaan et al. 2003)
${{\rm d}n}/{{\rm d}log_{10}M_{\rm HI}}$ in units of ${\rm Mpc}^{-3}$. 
The number of sources viewed assuming that this mass 
function is constant in $z$ will be

\begin{equation}
\frac{{\rm d}N}{{\rm d}z}=\int_{M_{\rm HI}(z)}^{\infty} 
\frac{{\rm d}n}{{\rm d}log_{10}M_{\rm HI}}
\frac{{\rm d}V}{{\rm d}z} {\rm d}log_{10}M_{\rm HI},
\end{equation}

\noindent where ${M_{\rm HI}(z)}$ is the limiting mass that can be detected; this quantity is
plotted in Fig. 1 in which the gain in sensitivity at high redshifts from a `tiled' survey is made clear.

\begin{figure}
\begin{center}
\centerline{\psfig{file=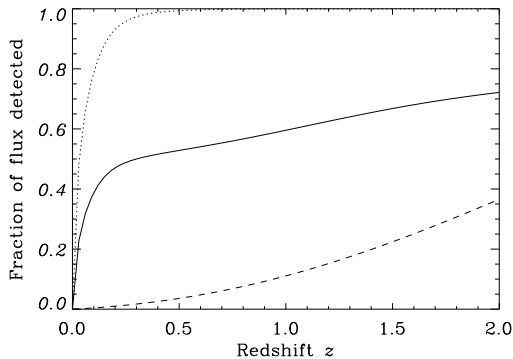,width=7.5cm,angle=0,clip=}}
\caption{The fraction of flux detected from an extended model
galaxy by an SKA-like interferometer as a function of redshift for different 
array distributions. 
The dashed assumes a scale-free array,
the dotted line assumes an array with just  
a core, and the solid line assumes the composite array as described in Sec. 2.6.}
\label{mass_limit}
\end{center}
\end{figure}

This assumes, however, that the HI mass function is constant throughout all redshifts 
which is clearly very unlikely to be correct as neutral Hydrogen is constantly being used in 
star formation, ionised in processes such as supernovae explosions as well 
as being created in processes like cooling flows (e.g. White \& Frenk 1991). 
We therefore have to try 
and see how this HI mass function evolves with redshift to have a clearer
idea of what number density will be accessible by surveys with next-generation radio telescopes. 
This is the purpose of Sec.3 where we assume that the HI observed traces collapsed 
dark-matter halos.

\subsection{Source visibility for an interferometer}

In the previous sections we have assumed that a source in the sky has a flux 
that will be detected perfectly by the radio telescope. This is not the case if the survey is
done with an interferometer. Since
the output of the correlator observing
an extended radio source on certain baselines does 
not recover the total flux density. 
We only recover the `correlated flux density' which is the modulus
of the complex visibility of each baseline.

Here we estimate the signal that we lose if we
do such surveys with an interferometer. We will assume that our average galaxy has 
a physical radius $R_0=15 \,{\rm kpc}$ at redshift 0. 
We assume that the dark-matter mass of a typical galaxy halo changes as 
$M_{\rm DM}^{\star}(z)$ according to the hierarchical growth of halos (Press \& Schechter 1978)
and that the characteristic density $\rho$ of a halo changes as 
$\rho \propto (1+z)^3$; i.e. that the density of the collapsed galaxy changes 
in the same way as the background density of dark matter. 
In the standard picture of disk formation (e.g. Peebles 1969; Fall \& Efstathiou 1980)
we expect the disk of a galaxy to be a factor of $\lambda$ smaller than the 
radius of the dark-matter halo, where $\lambda$ is the spin parameter and can be taken 
as a constant (Efstathiou \& Jones 1979).
Now given that $M_{\rm DM}(z)\simeq\rho(z) R^3(z)/R_0^3$ 
we can approximate the proper radius of an average galaxy as

\begin{equation}
R(z) \simeq 
R_0\left(\frac{M_{\rm DM}^{\star}(z)}{M_{\rm DM}^{\star}(0)}\right)^{1/3}\frac{1}{1+z},
\end{equation} 

\noindent where our choice of $M_{\rm DM}^{\star}(z) \propto (1+z)^{-3}$ 
is explained in Sec. 3.2. This will
impose an evolution of the characteristic sizes of disks with redshift that is 
proportional to $(1+z)^{-2}$; this is in rough agreement with observations of disk sizes from the
Hubble Deep Field (e.g. Poli et al. 1999; Giallongo et al. 2000). 
Other prescriptions for the evolution of the disk sizes could have 
been used (e.g. Ferguson et al. 2003) but this would not have made a large difference 
to our calculations. We use a characteristic size scale for an HI disk in a galaxy
at the break of the HI mass function at 
redshift $z=0$ of $R_0$ = 15 kpc (Salpeter \& Hoffman 1996).

We will assume that the galaxy has a surface brightness that corresponds to a 2-D Gaussian 
profile with angular spread equal to ${R(z)}/{D_A(z)}$ in radians, where $D_A(z)$ 
is the angular diameter distance to the galaxy. The Fourier transform of 
this Gaussian will be the complex visibility of each baseline. The spread 
of this Fourier transform is 

\begin{equation}
D_{\rm baseline} \sim \frac{c}{\nu_{12}} (1+z) D_A(z) \frac{1}{\pi R(z)},
\end{equation}

\noindent
so each antenna, when correlated with other antennas,
will only be sensitive to a fraction of the total flux of the source, 
depending on the baseline.

Obviously we need to assume a certain configuration for the antennas.
We adopt a composite array (see Jones 2004)
in which the interferometer will be configured with 20\% of the
collecting area inside a diameter of $\sim$ 1 km, 50\% of the collecting area within a diameter
of $\sim$ 5 km, 75\% of the collecting area within a diameter of $\sim$ 150 km and the final 25\% spread over
a diameter of 3000 km in a scale-free configuration (Conway 1998). 

\begin{figure}
\begin{center}
\centerline{\psfig{file=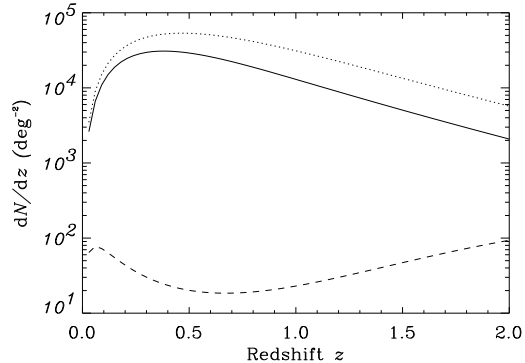,width=7.5cm,angle=0,clip=}}
\caption{Differential number density ${\rm d} N / {\rm d}z$ of sources per square degree
if we assume a fixed SKA-like sensitivity, a 4-hour integration time, 
a 10$\sigma$ detection limit and no evolution of the HI mass function (although 
galaxy radii are assumed to change with redshift in the way described by Eqn.11). Three
choices of antenna distributions are considered: 
(i) a simple scale-free array with resolution
0.1 arcsec at 1.4~GHz (Conway 1998) (dashed line); 
(ii) a core only (dotted line);
(iii) the composite array described in Sec. 2.6 (solid line).
Performing HI surveys with an array having a large fraction of the collecting area in short baselines 
is vital as many sources are resolved out by the longer baselines.}
\label{mass_limit}
\end{center}
\end{figure}

We illustrate how much the configuration affects the fraction of the
flux detected by plotting the number of sources seen with different configurations (Figs 2 \& 3). 
The fraction of flux detected from sources improves rapidly as the source 
moves to larger redshift
and has smaller angular size, and is therefore 
visible to a greater fraction of the interferometer baselines.
As seen in Fig.3 it is vital that a large fraction of the collecting area
is in short baselines so that we do not `resolve out' extended sources and 
miss most of the HI galaxies in a survey.

\subsection{Survey completeness limit}

If we perform a survey and take data out to redshift $z \sim 1.5$ we would need to cover 
frequencies from 1.4 GHz down to 560 MHz; with channels of width 30 km/s, we would have 
data in roughly
10000 channels. A survey over half of the sky at a resolution corresponding to
$ \simeq 1$ 
arcsec would have about 20000 square degrees, so we therefore would have around $10^{11}$
pixels in the sky. Our survey would therefore have around $10^{15}$ pixels in 3-D. 
We would therefore require a 8-$\sigma$ catalogue in order not to have any spurious 
sources if the noise was perfectly Gaussian.

Nevertheless, we expect to find sources of noise that will not be so well 
behaved. We therefore adopt a signal-to-noise level of 10$\sigma$ for
our calculations. In this way, we can be reasonably confident that our catalogue 
will be largely free of spurious sources.

We note that the HIPASS BCG (Bright Galaxy Catalogue;
Ryan-Weber et al. 2002) is a 9$\sigma$ catalogue,
and the number of pixels, in 3-D, in their survey is only $\sim 10^{8}$ given that they 
perform a survey out to 13000 km/s, have a resolution of around 13 km/s and 
survey half the sky with a resolution of order 15 arcmin. We therefore
conclude that depending on the sources of systematic 
errors, it might be necessary in a real survey to 
have a much higher detection level if we want the catalogue to be 
free of spurious sources (possibly higher than 10$\sigma$). We stress that 
such a change would not affect our results significantly.

\section{Possible evolution of the HI mass function}

\subsection{Information from damped-Ly$\alpha$ systems}

The number density of HI sources we will be able to detect will be a direct 
function of the evolution of the mass function of neutral Hydrogen. If we are able to 
construct a reasonable model that would predict this mass function at high redshift 
we would be able to have a good estimate of the number of sources that will be 
detected. 

If we perform an HI survey out to cosmological redshifts we will be 
sensitive mainly to objects near the break of the mass function. In the following sections 
we use scaling relations according to the properties of such galaxies
(i.e. we assume that the population at all redshifts have such properties). 
We will estimate the HI mass function at high redshifts 
but we are not too concerned if the low-mass objects are not correctly described.

In order to calculate the HI mass function at 
higher redshifts, we use damped-Ly$\alpha$ results as a probe. The current sensitivity of 
radio telescopes limit the detection of HI in emission to only $z\simeq 
0.2$ (Zwaan et al. 2001), so we can only probe high-$z$ objects 
in absorption. The damped-Ly$\alpha$ systems 
give us a distribution of the number density of objects as a function of the column 
density as well as the total $\Omega_{\rm HI}(z)$ at each redshift. One might think that
by looking at the distribution of column densities at high 
redshift we could find the distribution
in mass at high z. There is indeed a correlation between high column density and
high mass but the scatter is extremely large (Ryan-Weber et al. 2003)
because small clouds with not much HI can have 
lines of sight passing through their dense cores
giving a high column density and HI-rich galaxies can have lines of sight passing through
their low-column-density regions. We cannot obtain directly a mass function from
the distributions in column density, but we argue that the highest column
densities are mainly in collapsed structures (see Sec.4), 
and they account for most of the mass of neutral 
Hydrogen (Peroux et al. 2001). The total density of 
neutral gas at high redshift is the vital information that damped-Ly$\alpha$ 
systems give us, and this, of course directly constrains the HI-mass-weighted area
under the HI mass function.

\subsection{Possible evolution of the HI mass function}

We will assume that the mass function of HI can be 
described as a Schechter function at higher redshifts and that all
the Hydrogen seen in damped-Ly$\alpha$ 
systems is in collapsed halos at high redshift (see Sec.4). 
It is not known whether
this assumption is valid at higher redshifts, but we are confident that this is a reasonable 
approximation at least up to the redshifts in which 
we are interested. We therefore have three 
parameters to determine at every redshift: the normalisation $\theta^{\star}$, 
the faint-end slope $\alpha$ and the break of the mass function $M_{\rm HI}^{\star}$. 

All our calculations use the redshift-zero HI mass function measured by the HIPASS team 
(Zwaan et al. 2003). In their paper they parametrise the HI mass function by a 
Schechter function, namely

\begin{equation}
\frac {{\rm d}n}{{\rm d}(M_{\rm HI}/M_{\rm HI}^{\star})}=\theta^{\star} 
\left(\frac{M_{\rm HI}}{M_{\rm HI}^{\star}}\right)^{\alpha} 
exp\left(-\frac{M_{\rm HI}}{M_{\rm HI}^{\star}}\right) , 
\end{equation}

\noindent where $\alpha=-1.3$, $M_{\rm HI}^{\star}=10^{9.48}M_{\odot}$ 
and $\theta^{\star}=0.025$ ${\rm Mpc}^{-3}$ (Zwaan et al. 2003).
We assume that the faint-end 
slope will not play a big r\^{o}le in determining the number of sources to be seen. We 
keep this value constant and equal to the value ($\alpha=-1.3$) seen at $z=0$. 

The value of the normalisation is set according to the amount of gas found at
redshift $z$ following the results from damped-Ly$\alpha$ observations (Peroux et al. 2001); 
we assume that
the neutral gas in these damped-Ly$\alpha$ clouds is in collapsed objects.
The integral of the HI mass times the HI 
mass function equals the total density of neutral gas 
and is given by damped-Ly$\alpha$ results (see Eqn.14). 
Nevertheless, recent results show that if we could select quasars in the radio we might 
infer a higher amount of HI because there might be an
obscuration selection effect in finding neutral gas via damped systems selected in the optical 
(Ellison et al. 2001). We therefore multiply the Peroux et al. (2001) 
results by a small factor 
(namely 1.5; Ellison et al. 2001) to account for this potential selection effect.
We then fit a 
function to force the normalisation of our HI mass 
function to account for the amount of gas that must be present at that redshift; this is done
by forcing the integral under the HI mass function (weighted by $M_{\rm HI}$) to equal
that measured by the damped-Ly$\alpha$ results. 
We point out here that we have ignored the weak constraints on
$\Omega_{\rm HI}$ from Rao \& Turnshek (2000) who
predicted a larger amount of neutral gas at $z < 1.65$ on the basis of objects 
selected via MgII absorption and followed up with the HST to obtain damped-Ly$\alpha$ measurements;
their constraints are uncertain because of small number statistics, but also
because there are numerous systematic effects that may not yet have been completely 
understood (Rao \& Turnshek 2000).
From Peroux et al. (2001) we have

\begin{equation}
\Omega_{\rm HI}(z)
=\frac{1}{\rho_c}\int_{-\infty}^\infty 
M_{\rm HI}\frac{{\rm d}n}{{\rm d}log_{10}M_{\rm HI}}(z)\, {\rm d}log_{10}M_{\rm HI} ,
\end{equation}

\noindent which leads to

\begin{equation} 
\Omega_{\rm HI}(z)=\theta^{\star}(z)\Gamma(2+\alpha) M^{\star}_{\rm HI}(z)/{\rho_c},
\end{equation}

\noindent where $\Gamma$ is the gamma function 
and $\rho_c$ is the critical density of the Universe; we have used 
the Schechter function as the form of the HI mass function.

The only thing left to choose is how the break of this mass 
function will evolve in cosmic time. 
This is not well 
constrained by current data and can make 
a significant difference to our results. If we adopt the standard hierarchical 
formation scenario, where 
smaller objects merge to produce bigger objects, one would naively 
expect that this break would 
shift to lower masses at higher redshifts. Nevertheless, the problem is not so simple. 
There are many other processes that may lead to the opposite result. 
For example, if star formation is more efficient in Hydrogen-rich objects 
then they will tend to form stars at a higher rate than their lower-mass 
counterparts. In this case, if star formation is the main process
the break could shift 
towards higher HI masses at higher redshifts.

\begin{figure}
\begin{center}
\centerline{\psfig{file=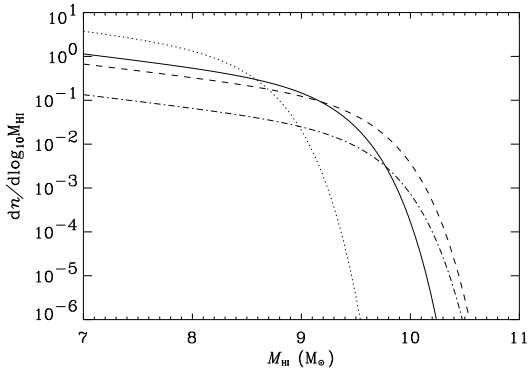,width=7.5cm,angle=0,clip=}}
\caption{The HI mass function plotted at redshift $z=2$ 
for Models A (dashed line), 
B (dotted line) and C (solid line). We also plot the
a Schechter-function fit to the measured
HI mass function at $z=0$ (dot-dashed line). Measurements of the total
density of HI in the high-$z$ universe from observations of
damped-Ly$\alpha$ systems (Sec.3.1) fix the area, in an HI-mass-weighted sense, under the curves 
for Models A, B and C, but the existing observational data are 
insufficient to establish the evolution of the 
location of the break in the HI mass function.}
\label{mass_limit}
\end{center}
\end{figure}

A complete theory of galaxy formation would be able to give us the answer to this problem. 
We do not however possess such a theory. We therefore will make 
three extreme choices for the change 
in the break. In Model A we consider a break that remains constant throughout all redshifts. 
In Model B we consider the case where $M_{\rm HI} \propto M_{\rm DM}$:
in this choice we are 
assuming that the HI follows a hierarchical formation in the same way that
dark matter clusters. In this Model B we 
use the fact that the Press-Schechter mass function (Press \& Schechter 1974) 
is roughly self-similar in 
$\nu={\delta_c}/[{\sigma(M) D(z)}]$ (Jenkins et al. 2001) where $\sigma^2(M)$ is the variance 
of the density field (at z = 0) smoothed
over a cosmic volume corresponding to mass $M$, $D(z)$ 
is the linear growth factor and $\delta_c \simeq 1.67$ is the linear theory 
threshold for collapse (Lokas \& Hoffman 2001). If we consider a 
change such that $\sigma(M) {\rm D}(z)$ 
remains constant we can see how a characteristic mass of dark matter changes with redshift. 
In fact $\sigma(M)\simeq M^{-(n+3)/6}$ where $n$ is the 
slope of the power spectrum at the scales of interest. 
We choose $n\simeq -1$ since this is appropriate for galaxy scales (Peacock 1999; p. 499). 
We can roughly say that 
$M_{\rm DM}\propto {\rm D}^3(z) \simeq (1+z)^{-3}$ in this model. With this choice,
the scaling of the average radius of a disk with redshift according to Eqn.11 would be roughly
$R(z) \simeq R_0 (1+z)^{-2}$, a scaling of galaxy sizes with redshift that is often simply 
assumed (e.g. Silk \& Bouwens 1999). 

The assumption that 
$M_{\rm HI} \propto M_{\rm DM}$ is not the most physically reasonable 
assumption we could make. In fact Model B is an extreme case, the function of which is 
only to provide a limit to our predictions. 
The main reason for this is that this model neglects the effects of star formation
over a range of cosmic epochs. However, we also know that this assumption 
must break down in high-mass halos (e.g. rich clusters)
where the ratio of $M_{\rm HI}$ to $M_{\rm DM}$ is much lower than in galaxies
(Battye, Davis \& Weller 2004). This is due, at least in part, to the long-established fact 
that the mass-to-light ratio is larger in the most massive halos due to long cooling timescales,
but it may also reflect a reduction in neutral content once galaxies become sub-halos
of a larger dark-matter halo (e.g. Zwaan et al. 2001). In extrapolating correctly to high redshift we would need to account for
times when these sub-halos were distinct halos, this would make the location of the
break of the HI mass function move to lower masses less rapidly than Model B.

In Model C we consider the case where 
$M_{\rm baryons} \propto M_{\rm DM}$. Once more,
this neglects the sub-halo problem but it does attempt
to take account of the star formation.
We assume that outflows will
reduce the fraction of baryons in a galaxy but accretion will bring the 
fraction back close to the nucleosynthesis value
or some relatively fixed value of it (Silk 2003); a steady
state between these two processes would bring the ratio
of baryons to dark matter to a constant value. 
In this case we can say that 
$M_{\rm HI} \propto [{\Omega_{\rm HI}(z)}/({\Omega_{\rm stars}(z)+\Omega_{\rm HI}(z)+\Omega_{\rm H_2}(z)})] M_{\rm DM}$,
where $\Omega_{\rm stars}(z)$ is given by

\begin{equation}
\Omega_{\rm stars}(z)=\frac{1}{\rho_c}\int_z^\infty 
\frac{{\dot{\rho}}_{\rm stars}(z)}{H(z)(1+z)}{\rm d}z,
\end{equation}

\noindent and where the star formation rate ${\dot{\rho}}_{\rm stars}(z)$ 
as a function of cosmic 
time is taken from Choudhury \& Padmanabhan (2002) and corrected to the cosmology we use here
[$H(z)$ is the Hubble Constant at redshift $z$]. 
The choice of $\Omega_{\rm H_2}$ is more complex. From Young \& Knezek (1989) 
we know that the ratio of molecular Hydrogen (${\rm H_2}$)
to neutral Hydrogen (HI) is a function 
of the galaxy type varying by a factor $\sim 20$ depending on type.
However Young \& Knezek argued that the largest ratios ($M_{\rm H_2}/M_{\rm HI}\sim 4$) 
are found in the most massive galaxies with deepest 
potential wells that correspond to dense 
stellar cores with low spin temperatures. 
Less massive galaxies have a smaller proportion of molecular hydrogen
($M_{\rm H_2}/M_{\rm HI}\sim 0.2$) and have less dense cores with high spin temperature. 
We cannot implement this in a 
completely consistent way in our Model C, so we decide to choose a ratio so that 
$\Omega_{\rm HI}(z)=\Omega_{\rm H_2}(z)$ which is consistent with the local 
baryon budget (Fukugita, Hogan \& Peebles 1998). 
We also 
neglect the ionised fraction of gas in a galaxy i.e. ($\Omega_{\rm HII} \simeq 0$)
which is a reasonable assumption for collapsed galaxy-sized halos (Fukugita et al. 1998). 
In Model C the break 
of the HI mass function will still shift toward smaller HI 
masses at higher redshifts, but 
more slowly than Model B as it takes account the fact that 
galaxies at higher redshifts have more gas and less stars.

\subsection{Limits on the number count}

We plot in Fig.5 the differential number 
density ${\rm d}N/{\rm d}z$ of objects seen, per ${\rm deg}^2$, 
by a future SKA-like telescope for our Models A, B and C; 
we also plot the number density with a HI mass function that does not
evolve with redshift. We can see that the evidence from damped-Ly$\alpha$
systems for more HI in collapsed objects (Sec. 3.1) predicts a larger number density of 
HI-emitting objects. 

\begin{figure}
\begin{center}
\centerline{\psfig{file=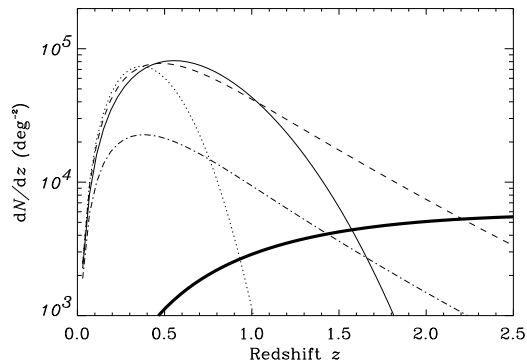,width=7.5cm,angle=0,clip=}}
\caption{The differential number density (${\rm d}N/{\rm d}z$)
per ${\rm deg}^2$ of objects in an SKA-like survey with a signal-to-noise 
detection level of 10, and an integration time of 4 hours 
in a tiled survey for our 
three evolution models [Model A (dashed), B (dotted) and C (solid)] 
plus a no-evolution model (dot-dashed).
The integral under these curves represent the total number of HI emitting objects
per ${\rm deg}^2$ which are $8.0\,10^4$, $6.5\,10^4$, $2.0\,10^4$ and $3.5\,10^4$,
for Models A, B, C and the no-evolution model respectively. We note
that we assume more neutral gas at higher redshifts in the three
evolution models, so the number of HI sources in these models increases:
depending on our the choice of break in the HI mass function, 
the models probe out to different redshifts. The 
thick line defines the number density of objects needed at high redshift in order
for us to be cosmic variance limited when reconstructing the galaxy power spectrum (see Sec. 5.2).}
\label{mass_limit}
\end{center}
\end{figure}

We also see that considering a different break in the HI mass function 
makes some difference to the amount of objects seen. Even though we 
tried to choose very different possible alternatives we see that there is not a huge 
difference in the total number of HI sources. There are two reasons for this: first, 
if we consider a change in the position of the break of the HI mass function
and still consider that the total HI 
mass is the same at a given redshift the model with a lower break will have a lot more 
low-mass objects to account for the same mass at that redshift. If we have a survey 
that has enough sensitivity to reach those 
masses we will see those objects. Second, the change 
in the break becomes significant only at higher redshifts and these redshifts are probed 
at longer integration times. We can therefore say that for low integration times
the total number of HI sources is uncertain to a factor of only $\sim2$ (see Fig.5).

Although different breaks in the HI mass function 
give roughly the same amount of sources they
yield surveys probing significantly different 
volumes in space. We will define the depth of our survey ($z_{\max}$) carefully
in Sec.5.2, but we can already see (Fig.5) that, for a 
4-hour integration time, 
Model A probes redshifts as high as 2.1, whereas Model B only probes redshifts 
as high as 1.0. Model C, chosen to have a break in between the breaks for Models A and B, 
probes out to redshift 1.5.

Our preferred model in the next sections will be Model C. In Sec.6 we will 
quantify how much longer or shorter a survey will take if the HI mass function 
is closer to Models A or B (or if we have an 
evolution that mimics the no-evolution scenario) instead of Model C 
in order to get the same cosmological 
constraints on $w$.

\section {Other constraints on the evolution of the HI mass function}

In Sec. 3 we tried to see how the HI 
mass function could change with redshift.
We stress here that we are only concerned whether the objects near the
break of this mass function are well described as they are the ones that will dominate future
surveys at the redshifts of interest. We have assume that an average galaxy 
at $z=0$ has a radius 
of $\sim$15 kpc and a circular velocity of 200 km/s. We have also
assumed scaling relations that describe
their behaviour at higher redshift. We have assumed no scatter on these relations. 

To obtain an expression for the normalisation of the mass function we have assumed 
in Sec.3.2 that damped-Ly$\alpha$ systems are collapsed objects 
with radii of the same size as galaxies at the redshifts they are observed. 
If we assume that the path length through the absorber is roughly $l \simeq N_{\rm HI}/n_{\rm HI}$ 
where $N_{\rm HI}$ is the column density and $n_{\rm HI}$ is the HI number density and
$f_c$ to be the ratio of the density in baryons in a collapsed halo to the universal 
density of baryons (e.g. from nucleosynthesis), we have
$l/{\rm kpc}=109 \, (N_{\rm HI}/10^{23}{\rm m}^{-2}) \, (f_c/180)^{-1}(1+z)^{-3}$ 
(Peacock 1999; page 365) 
and this means that the most extreme damped systems ($N_{\rm HI}\sim 10^{24} \, {\rm m^{-2}}$) 
are consistent with having the dimensions of galactic disks: at $z=1$ the 
damped systems would have 
radii of around 1-10 kpc. 
This is not the case for much lower column density systems. Nevertheless the 
total mass obtained by integrating the damped-Ly$\alpha$ systems
over column density is mainly due
to high column density objects (Peroux et al. 2001), 
i.e. those that are most likely to be collapsed and to have the sizes of galactic disks.

When a blind HI survey is made in the local Universe, 
it is reported that the essentially all of the galaxies found have
optical counterparts (Ryan-Weber et al. 2002). 
If a large fraction of the gas were in non-collapsed objects at 
higher redshifts then this would
reduce significantly the number of detections by a future survey. There have been searches 
at low redshifts for 
large clouds of gas with high mass and very low column 
density (Minchin et al. 2003); these searches have 
found no large clouds with low column density, and have also found that all collapsed galaxies
detected in HI have an average column density $N_{\rm HI}\sim10^{23}$ to
$10^{25} \, {\rm m^{-2}}$. In Ryan-Weber et al. (2002),
a sample of 34 galaxies from
this HIPASS survey were observed at higher spatial resolution
so that the column density could be computed at each point of each galaxy. 
The resulting column density distribution is similar to those in
to damped-Ly$\alpha$ results (Peroux et al. 2001), allowing for an increase in 
normalisation for the reasons outlined in Sec.3., but in this case we know
that the galaxies in HIPASS are in collapsed objects. 
This is strong evidence that the HI found in blind surveys traces the dark-matter
potential wells in a similar way to the baryons, and this suggests 
that the HI found at high redshift
is likely to be in collapsed objects such as young galaxies. 

In our Model C we have assumed that the mass in baryons in a galaxy follows
the dark-matter mass. Observationally, there is compelling evidence that this is
the case if we look at the Tully-Fisher relation in spirals. McGaugh et al. (2000)
have shown the Tully-Fisher relation obtained using only the stellar 
component of spiral galaxies has a break at around 90 km/s but this break disappears if, 
instead of considering just the stellar population, they use the total 
baryonic mass of the galaxy composed of gas plus stars. They get a very good fit
for both ends of their data, i.e. for low and high circular velocities. 
This baryonic Tully-Fisher relation has the following implication: the mass of gas
plus stars is directly proportional to the mass in dark matter, and any other
dark component is unlikely to be important as it would introduce too much scatter. 

If we look at the blue luminosity function of galaxies at redshift $z=1$, 
we can use scaling relations in order to estimate whether it is 
consistent with Model C. We crudely assume that the
blue luminosity density is proportional to the star formation rate at each epoch. 
Therefore, the ratio of the breaks of the HI mass function and the blue luminosity 
function should be in proportion to the ratio of $\Omega_{\rm HI}$ to the star
formation rate. Given that we have an estimate of the star formation rate 
and the density of neutral Hydrogen at $z=1$ (Sec. 3.2),
we can infer the position of the break of the $z=1$ 
HI mass function by knowing the
position of the break of the $z=1$ luminosity function of blue galaxies: 
this has been taken from COMBO-17 
(Chris Wolf; private communication based on the data from Wolf et al. 2003). 
We plot and compare this model with our Models A, B and C. We can see from Fig.6 that Models A and C are 
clearly preferred by this comparison. In fact at redshift $z=1$
our Models A and C give essentially the same prediction for the HI mass function,
as can be inferred from the similar ${\rm d}N/{\rm d}z$ for these 
two models over the redshift range $0 \ltsimeq z \ltsimeq 1$.

\begin{figure}
\begin{center}
\centerline{\psfig{file=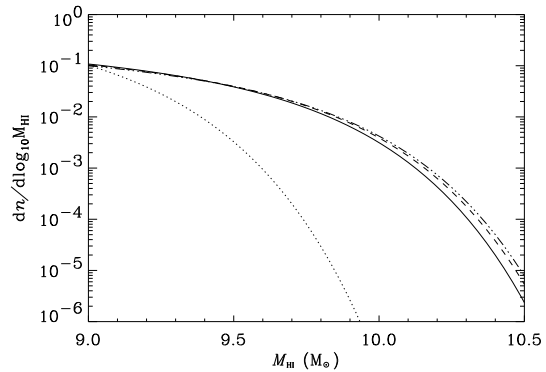,width=7.5cm,angle=0,clip=}}
\caption{The HI mass function plotted at redshift $z=1$ for Models A (dashed line), 
B (dotted line) and C (solid line). We compare these models with
the HI mass function that is inferred from the optical luminosity function of blue galaxies
as described in Sec.4
(triple-dot-dashed). The curves for Model A, Model C and the 
model inferred from the blue luminosity function are very 
similar and almost indistinguishable from each other.
This indicates that the 
HI mass function is likely to be closer to Models A or C than to Model B.}
\label{mass_limit}
\end{center}
\end{figure}

In our model we assume that the average dark-matter halo in a galaxy at the 
break of the HI mass function is smaller at higher redshifts according to hierarchical 
processes but, on the other hand, these galaxies have less stars and are more gaseous. 
So, if we take galaxies at $z \simeq 2$, their average luminosity will
be roughly proportional to the stars in that galaxy which will be approximately 
$[{\Omega_{\rm stars}(z)}/({\Omega_{\rm stars}(z)+2\Omega_{\rm HI}(z)})] L^{\star}$
so, at $z=2$, these galaxies would be, assuming no scatter,
50-times fainter optically than 
$L^{\star}$ (i.e. objects at the break of the optical Schechter luminosity function). 
Deep near-infrared imaging with the HST typically fails to detect stellar counterparts to
damped-Ly$\alpha$ objects (Warren et al. 2001) to limits which are consistent with 
our rough calculation based on Model C.

It is also well known that it is very hard to find known 
damped-Ly$\alpha$ systems in HI absorption in the radio. This is 
also in agreement with Model C as, at high redshifts, 
this predicts smaller gas rich-clouds with shallower gravitational 
potentials and consisting mainly of a warm neutral medium that 
has a high spin temperature (Kanekar \& Chengalur 2003). 
Because of the high spin temperature it 
will be hard to find these objects 
in absorption in the radio until we have instruments 
with the sensitivity of the SKA.

Finally, there has been searches for H$\alpha$ lines associated with star 
formation in damped-Ly$\alpha$
systems, but typically no line flux is found. This can impose
(assuming a negligible dust obscuration) a maximum star formation 
rate (SFR) which would correspond 
to 11.4-36.7$ \, M_{\odot} {\rm yr}^{-1} $ for the studied objects (Bunker et al. 1999). 
Since our break for the HI
mass function moves towards the left for our preferred Model C, the star formation 
rate in high-$z$ objects would be 
lower than in galaxies at the break of the low-$z$ HI mass
function. We therefore
would not expect any line flux to be detected is such searches
as the SFR is likely to be too low to produce a robust detection, so these results are
also in agreement 
with Model C. Nevertheless, if we do have a star formation 
$\sim 10 \, M_{\odot} {\rm yr}^{-1} $, we would have an associated continuum emission
in the radio of $\sim 1 \, \mu \rm Jy$ at redshift $z=2$ (Condon 1992), 
and we will trivially be able to detect this level of emission with SKA-like instruments (Eqn.2).
In fact the continuum emission 
is only a factor of 15 fainter than the corresponding line emission. 
Good band-pass calibration will therefore be 
essential for the line detection of star-forming
galaxies, particularly those that may, because of inevitable scatter,
have a continuum flux of the same order of magnitude as the 21-cm line emission.

We have to stress here that Model C is basically taking the known population
of HI-emitting objects at $z=0$ and trying to see what they would look 
like at higher redshift. It is likely though that there will be another population
of HI-emitting objects which would be the sub-units of large elliptical galaxies today.
These sub-units would probably have been HI-rich in the
high-$z$ universe, before they merged to form large elliptical galaxy
and before they became part of large dark-matter halos such as groups and clusters. 
We think, nevertheless, that these objects would still be rare at redshifts around
$z\simeq1.5$. First, given the colour information on elliptical 
galaxies found in observations (Bower, Lucey \& Ellis 1992), it is argued that most of the stars
in giant elliptical galaxies
must have been formed at redshifts higher 
than 1.5 to explain their consistently red colours. 
Second, a population of extremely red objects (EROs) is now well established
at $z\simeq1.5$ (e.g. Daddi, Cimatti \& Renzini 2000) which look to be progenitors
of nearby ellipticals, and which appear to be red because of the old stellar
populations as was first demonstrated for $z\simeq1.5$ radio galaxies (Dunlop et al. 1996).
Third, if such a population exists it is likely that they would appear as large
spiral galaxies at $z\simeq1.5$ but observationally (Wolfe et al. 1985) only one 
such example has been detected as a damped-Ly$\alpha$ system at 
$z\simeq2$. Most damped-Ly$\alpha$ objects are consistent 
with being smaller galaxies with a warm neutral medium (e.g. Kanekar \& Chengalur 2003)
which would be the precursors of the spiral population today.
We neglect the elliptical population but caution that it may
start to contribute significantly to the HI population at redshift $z\sim2$ (Sec.6).

We have emphasised here how the break of the HI mass function is likely to
change with redshift but this change is not directly 
constrained by current data. However, we have discussed various
observational constraints in this Section which are in good agreement 
with our Model C. If the break of the HI mass function were to stay 
at very high masses at 
higher redshifts then we would expect brighter galaxies and 
stronger star-formation lines associated
with most damped-Ly$\alpha$ objects.
On the other hand, if the HI break shifts towards very low masses at higher redshifts 
it is hard to find consistency with the blue luminosity function at $z=1$ (Fig.6).
Finally we note that our Model C is in good agreement out to $z\sim 1.5$ with the 
predictions of semi-analytic models; see Rawlings et al. (2004) for a comparison
with models from Cole et al. (1994) and Benson et al. (2003).

\section {Probing dark energy via baryonic oscillations}

It has been proposed in the literature by several authors
(e.g. Blake \& Glazebrook 2003; Hu \& Haiman 2003; Seo \& Eisenstein 2003) that measuring the 
baryonic oscillations in the galaxy power spectrum allows a clean method of probing
properties of dark energy which could be performed 
provided enough cosmic volume and enough tracers of this volume (e.g. galaxies) are available.

There are two sources of error in such 
a power spectrum measurement. The first of these is sample 
or cosmic variance which is linked to the fact that 
the number of independent spatial modes that we can measure in a
given cosmic volume is finite. This error is inversely proportional to the 
square root of the cosmic volume
covered by a survey (see Eqn. 17). 
In the case of an HI survey with an SKA-like telescope this will be determined by the 
area of the sky we will be able to survey and the integration time we will spend 
in each square degree which will in turn, constrain the maximum redshift 
at which we will be able 
to detect HI sources (see Sec. 5.2 and Fig. 5). The other source of error is
shot noise due to the imperfect sampling of the fluctuations due to the finite number of 
tracers in the volume. The total fractional error in the power spectrum assuming the optimised weighting scheme of Feldman, Kaiser \& Peacock (1994) is:

\begin{equation}
\left(\frac{\sigma_P}{P}\right)^2=2
\frac{1}{4 \pi k^2 \Delta k}\frac{(2\pi)^3}{V_{\rm survey}}\left(\frac{1+nP}{nP}\right)^2,
\end{equation}

\noindent where $P$ is the value of the power spectrum at wavelength $k$
and $n$ is the number of sources per volume in our sample \footnotemark.

\footnotetext{We adopt the Fourier transform convention in which $nP$ has no units.} 

We would like to minimise the error we get from a P(k) measurement. This
would involve designing a survey that would have maximum volume provided that there is
enough
sources so that the shot noise is negligible compared to the error due to cosmic variance.
In order to have a negligible shot noise we would need $nP \gg 1$ (Eqn. 17) but this would 
involve getting an enormous amount of sources per volume. As is argued in 
detail by Seo \& Eisenstein (2003)
there is not a great gain in going from $nP \sim 3$ to $nP \to \infty$ so 
we assume that our errors will 
become large compared to cosmic variance only when $nP<3$.

\begin{figure}
\begin{center}
\centerline{\psfig{file=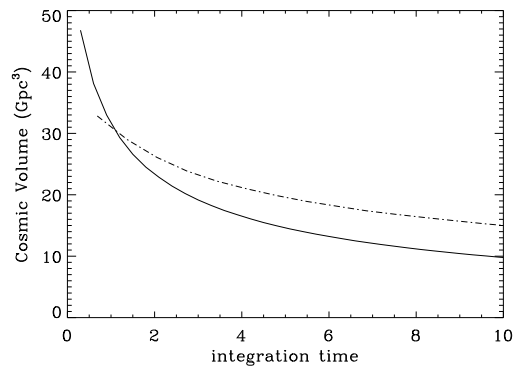,width=7.5cm,angle=0,clip=}}
\caption{The solid line shows 
the cosmic volume surveyed in one year for a 1 ${\rm deg}^2$-$FOV$ SKA-like instrument
at 10$\sigma$ detection level and with a $FOV$ scaling with frequency $\nu$ as $\nu^{-2}$;
this is plotted as a function of integration time per pointing. 
Note that we may choose to have a short integration time and 
cover a large fraction of the sky or have a deeper survey over a smaller area.
Note also that, assuming a dedicated telescope and negligible overheads,
if we choose an integration
time as low as $\sim0.3$ hours we can cover half of the sky in $\sim1$ yr.
It is clear that the optimal survey in terms of volume access is the shallower survey over 
all the sky available. The dot-dashed line is the volume weighted by the number of wiggles 
in the linear part of the power spectrum for the same survey. We stress that it has no 
physical value but that it reflects the fact that the volume at higher redshift is more 
useful for dark energy parameter constraints. We can see that even with the fudge factor 
that takes account for this, an all hemisphere survey is still the best option 
for an SKA-like instrument.
}
\label{mass_limit}
\end{center}
\end{figure}

Having $nP \sim 3$ means that ${N}/{V} \sim {3}/{P}$ so we have 
errors in the shot noise that will be small if
${{\rm d}N}/{{\rm d}z}>({3}/{P})({{\rm d}V}/{{\rm d}z})$: here ${{\rm d}N}/{{\rm d}z}$ 
is the differential number density of sources that we get from our
model, and ${{\rm d}V}/{{\rm d}z}$ is the usual comoving volume element. 
Given that, at redshifts of interest ($z\sim1.5$) 
the `wiggles' in the power spectrum will have been erased by
non-linear clustering for $k \gtsimeq 0.25 \, {\rm Mpc}^{-1}$ (see Blake \& Glazebrook 2003),
we take $P(0.14 \, {\rm Mpc}^{-1})$ for our calculations, i.e. the position of the first wiggles in the power spectrum. 
We have assumed that HI galaxies have a bias that increases slowly as a function of redshift roughly as g($z$), 
from a bias of $\sim 1$ at low redshift: the close association at low redshifts between
galaxies selected by their HI content and the normal `late-type'
galaxy population (e.g.\ Minchin et al. 2003) and the unit bias
of this population with respect to the dark matter (Peacock et al. 2001; Verde et al. 2002), 
suggest that this assumption is reasonable.
We then assume that the systematic rise in the bias cancels with the drop in the normalisation 
of $P(k)$. We estimate that measurements 
will be statistically useful until the differential number density drops below ${\rm d}N/{\rm d}z\sim$ 5000 
galaxies per ${\rm deg}^2$ (see Fig. 5). This defines the
maximum redshift we can probe with a given integration time.

\subsection{Optimal survey strategy}

The error on the dark-energy parameter $w$ will not depend only on the cosmic volume probed and the shot-noise
of the experiment. 
It depends also on the number of wiggles probed 
[which is a function of the redshifts surveyed given that non-linearities can dilute and erase these wiggles], 
represented by $n_w(z)$ 
and on the strength of the test which can be represented 
as the distortion of the wiggle length as a function of redshift 
[see Fig. 5 of Blake \& Glazebrook (2003)]. Note from 1/$k^2$ dependence of Eqn. 17 that high-$k$
wiggles are easier to measure provided that they have not been 
diluted or even erased by non-linearities.

First, we calculate the best shape for a survey so that we get the 
biggest volume in the smallest time. Would it be a shallow survey that would cover 
the whole sky or a deep survey that would cover
only a fraction of the sky? In order to determine this we compute the 
volume surveyed by multiplying the area covered in the sky by the effective volume 
covered by the data. From Tegmark (1997) we get

\begin{equation}
V_{\rm survey}=\frac{T_0}{t_1}\frac{FOV}{20000 \, {\rm deg}^{2}} \int_V \left(\frac{nP}{1+nP}\right)^2 {\rm d}V,
\end{equation}

\noindent where $T_0$ is the total time of the survey, and $t_0$ is the integration time 
spent per pointing, $t_1=t_0 / \beta$
where $\beta=min(1,{BW_{\rm SKA}}/{BW_{\rm survey}})$ and 
where $BW_{\rm SKA}$ stands for the bandwidth 
allowed given a particular realisation of an SKA-like instrument
and $BW_{\rm survey}$ represents the frequency range corresponding to HI redshifted
throughout the range of redshifts of the survey. In most SKA realisations 
$\beta$ is expected to be below one, but should not 
be an order of magnitude below one.

We choose a survey time of one year and a $FOV$ of 1 ${\rm deg}^2$,
and then set the optimum
integration time that will maximise $V$.
As we can see from Fig.7 we obtain the largest volumes when we perform a shallow survey 
across all the available sky. 

Nevertheless the survey with largest volume is not necessarily 
the best survey to probe dark energy. A survey at high redshift may have more wiggles 
and therefore provide a better constraint than a survey with a large volume at low 
redshift. In order to illustrate this we plot in Fig. 7 a volume weighted by 
$n_w(z)$. We stress here that this is only a toy model to illustrate that even though the test
may be more efficient at high redshifts. The sensitivities and $FOV$ of the SKA
are such that
a survey with the largest area is likely still to be the optimal survey. We expect to produce a more 
rigorous calculation of these effects.

If we look at Fig. 5 of Blake \& Glazebrook (2003) we can see that the baryonic oscillations 
test is considerably weaker below a redshift of 0.5. On the other hand we can see 
that even with this volume-weighted function in Fig. 7 the largest volume-weighted 
volume we obtain is for integration times that have the largest area available in the sky. 
All of these surveys probe redshifts larger than 0.5, in fact as we can see from Fig. 8 these 
surveys probes volumes at least as high as $z\sim 0.7-0.8$. A survey that is designed to probe dark 
energy with baryonic oscillations must have a considerable volume at redshifts larger than 0.5. 

We therefore conclude that the optimal way of probing Large Scale Structure 
with future radio surveys with a sensitivity comparable to that of an SKA 
will be to produce surveys that probe all the area available on the sky. We caution that this might not 
be the case 
if the telescope sensitivity is significantly lower.

\subsection{Surveys attainable by future radio telescopes}

The factor $nP/({1+nP})$ in Eqn. 18
will be very close to one for most redshifts where we have data 
and will fall sharply to zero where the HI starts to become too faint to be detected. 
Here we consider $nP/({1+nP})$ as a step-down function that becomes zero 
at $z_{\rm max}$. We define this maximum redshift $z_{\rm max}$ as the redshift where we
become shot-noise limited (i.e. $nP \sim 3$) which corresponds to 
${\rm d}N/{\rm d}z \sim 5000 \, {\rm deg^2}$. We therefore have

\begin{equation}
V_{\rm survey} \simeq 
\frac{T_0}{t_0} 
\frac{\beta FOV}{20000 \, {\rm deg^2}} \int_0^{z_{\rm max}(t_0)} \frac{{\rm d}V}{{\rm d}z}{\rm d}z,
\end{equation}

\noindent so the
value of $z_{max}$ will determine the maximum depth to which we will be able to
reproduce a galaxy power spectrum reliably.

\begin{figure}
\begin{center}
\centerline{\psfig{file=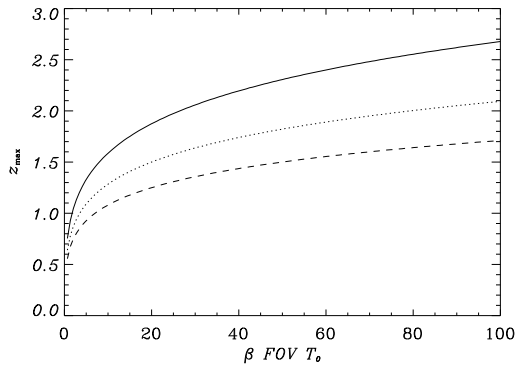,width=7.5cm,angle=0,clip=}}
\caption{Maximum redshift $z_{\rm max}$ 
probed by a survey as a function of telescope specifications and
duration $T_0$ of an SKA survey in units of yr. 
We assume that half the sky is observable so the integration time we 
choose trades off linearly with the $FOV$. 
A larger $FOV$ will allow us to spend longer in a patch of the
sky and therefore to probe to deeper redshifts.
The solid line is for a telescope with a field of view scaling with 
frequency $\nu$ as $\nu^{-2}$, 
the dashed lines is for surveys with telescopes with a constant field of view and 
the dotted line is for surveys with 
telescopes with a field of view scaling as $\nu^{-1}$. The three
curves have SKA-like sensitivity.  
}
\label{mass_limit}
\end{center}
\end{figure}

In Fig. 8 we plot the maximum redshift of a given survey with an SKA-like
instrument as a function of how the field of view
scales with $z$. 
There are four features of the future radio telescope 
that are vital in order for this survey to be optimal.
The first feature is the sensitivity of the instrument. 
As we can see from Fig. 8, with an SKA-like telescope
our gain in $z_{\rm max}$ starts dropping quickly as we start probing redshifts $\sim$ 1.5. 
We conclude that if redshifts of order 1.5 have to be 
reached in order to probe dark energy optimally
with this survey, then a full SKA-like-sensitivity is needed. 
As we will see in Sec. 5.3 this is indeed the
case and the full SKA will be needed to properly probe dark energy.

The second and third features are the $FOV$ and $\beta$, the 
useful bandwidth of the telescope, which
are essentially degenerate. We can clearly
see in Fig.9 that although SKA-like sensitivity is vital to get a large
cosmic volume, the factor $\beta \, FOV$ plays an equally important role. 
The fourth and final 
feature is the way the $FOV$ scales with frequency. We can see in Figs 8 and 9 the 
effect of this choice, and show the huge advantage that 
is gained if the SKA design can have field of view
scaling with frequency as $\nu^{-2}$.

The choice of these four features will indicate the 
depth of an eventual survey. For example
a telescope with $f=1$, $\beta \,FOV=1 ~ \rm deg^{2}$ and a field of view
scaling as $\nu^{-2}$ can survey 40 
${\rm Gpc}^3$ in one year. In ten years the same telescope will be able to survey 150 ${\rm Gpc}^3$
by probing deeper in redshift. this same 150 ${\rm Gpc}^{3}$ could be completed in just one year if
$\beta \,FOV= \, 10 \, \rm deg^{2}$.

We note here the angular resolution 
required by the baryonic oscillations method is much less than the angular
resolution expected of future radio interferometers like the SKA. 
The results from Fig. 9 are independent
of the resolution of the instrument because the baryonic `wiggles' probe very large angular 
scales, and confusion will not be a serious issue given the accurate redshifts available 
for objects in a low-resolution HI survey.
This method will therefore be able to be used
whether the instrument being used is an SKA-like array or the core of such an array.

\begin{figure}
\begin{center}
\centerline{\psfig{file=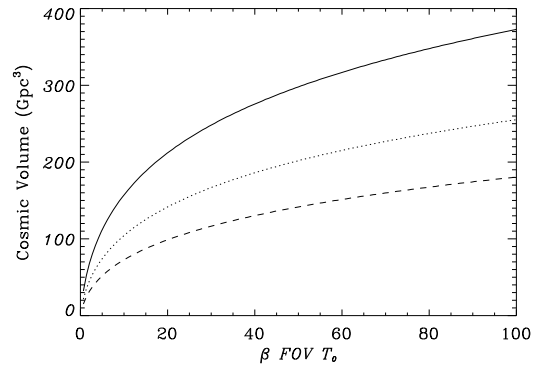,width=7.5cm,angle=0,clip=}}
\caption{Cosmic volume probed by a survey with a future radio telescope as a 
function of telescope specifications and length of survey at a 10$\sigma$ level of detection.
The solid line is for telescope with a field of view scaling with frequency 
$\nu$ as $\nu^{-2}$, 
the dashed lines is for surveys with telescopes with a constant field of view and 
the dotted line is for surveys with 
telescopes with a field of view scaling as $\nu^{-1}$. The three
curves have SKA-like sensitivity.
}
\label{mass_limit}
\end{center}
\end{figure}

\subsection{Constraints on dark energy}

Now, given that we can have a realistic idea of what type of 
survey future radio telescopes
may produce, we can relate these surveys to
an estimated error on the measurement of $w$. There are 
are likely to be other efforts in 
measuring the equation of state of dark energy 
using the baryonic oscillations method. As argued in 
Blake \& Glazebrook (2003), an optical survey would, in a one year, be 
able to measure $w$ to $\Delta w \simeq 0.1$, provided a spectrograph
that can take data on 3000 galaxies at a time is available on an 8-m 
optical telescope. In one year, a dedicated 8-m telescope with such an instrument
could cover a cosmic volume of $\sim 6\, {\rm Gpc}^3$
(a volume 6-times greater than that covered by the SDSS survey).
This KAOS project has been proposed (Barden 2003; Glazebrook 2003) 
and may produce results in the next decade. 

As we can see from Eqn.17, the error on the power spectrum scales as the 
$V_{\rm survey}^{-1/2}$, and this error would relate to the error
on the size scale of the wiggles which would in turn directly relate to an 
error on the parameter $w$. We therefore expect the error on $w$
to improve as $V_{\rm survey}^{1/2}$ as we cover more volume. 
At the likely rate of data collection, after a decade of results, the KAOS project is likely 
to produce a constraint roughly a factor $\sim f_{\rm sky}^{0.5}=\sqrt{10}$ 
better and therefore constrain $w$ 
down to $\Delta w \simeq 0.03$, surveying a volume\footnotemark of around 60 ${\rm Gpc}^3$.

\footnotetext{Note that optical surveys with KAOS cover sky area to similar volume depths at a similar rate to an SKA with
$FOV \, = \, 1 {\rm deg}^2$, but can never be fully dedicated to such a survey because of daylight and bright phases of moonlight.}

The volume surveyed in a one year survey with a dedicated radio telescope with $f\, = \,1$ and 
$\beta \, FOV \simeq 1 ~ \rm deg^{2}$ and a field of view scaling as $\nu^{-2}$ 
is around $40 \, {\rm Gpc}^3$. Given that it is
unlikely that an array such as the future SKA will be dedicated to a single project,
we conclude that if $\beta \, FOV \sim 1 ~ \rm deg^{2}$ for 
the SKA we would only be able, on the timescale of years 
to get constraints of around
$\Delta w \simeq 0.03$ (see Fig.9), comparable to those from KAOS.

We therefore argue that a future radio telescope with a small 
($\sim 1 \,{\rm deg^2}$) $FOV$
could constrain $w$ well, but similar constraints will already be available by the time
this telescope is operational. 
However, a data set with $z_{\rm max}\sim 1.5$ in a hemisphere,
which is achievable in one year provided $\beta \, FOV \gtsimeq 10$ (see Fig.8),
would get constraints 
of $\Delta w \simeq 0.01 \, (f_{\rm sky}/0.5)^{-0.5}$ 
given the extra volume available.
We would be able to improve considerably on these constraints 
in following years of survey by 
probing deeper in redshift and accruing more volume (see Fig.9).
If we parametrise the equation of state 
in the form $w=w_0+w_1z$ we would have $\Delta w_0 \simeq 0.035$ and 
$\Delta w_1 \simeq 0.1$ (Blake et al. 2004). Such a data set would be
ideal to probe the properties of dark energy and its evolution with redshift.

We conclude that for an SKA-like telescope, studies
of dark energy demand that the optimal telescope has
a field of view scaling with frequency $\nu$ as $\nu^{-2}$,
and an $\beta \,  FOV\gtsimeq 10 \, {\rm deg}^2$.

\section{Uncertainties and possible problems with future HI surveys}

\begin{table}
\begin{center}
\begin{tabular}{|l|c|c|c|}
\hline
Changes  & Change in time required to complete survey\\
\hline
$FOV$  & /$FOV$  \\
$FOV$ $\propto \nu^{-1}$ & x$\sim$2  \\
$FOV$ $\propto \nu^0$ & x$\sim$4  \\
Bandwidth & /$\beta$ \\
\hline
Model A & x$\sim8$ \\
Model B & /$\sim$2 \\
No evolution & x$\sim$1 \\
\hline
\end{tabular}
\end{center}
\caption{As mentioned in Sec. 5.3, with an `all hemisphere' survey covering redshifts 
0 to 1.5 we would be able to measure $w$ to $\Delta w \simeq 0.01$; 
this table illustrates how 
much longer/shorter it would take if we had chosen a different telescope design, 
(rather than a field of view scaling as $\nu^{-2}$) 
or a different model (rather than Model C) 
for the HI mass function. For example, a telescope with twice the $FOV$
would do the survey $\sim 2$-times 
faster, or if the real HI mass function where closer to Model A, 
the survey would take $\sim 8$-times longer.} 
\end{table}

In the previous sections we have assumed certain generic features of future radio surveys,
and we have also assumed
a certain evolution for the mass function of HI. With those two ingredients we 
have derived a 
number density of sources that would be accessible to us if we perform a survey 
of HI with a future radio telescope. In this section we 
relax some of the key 
assumptions, one by one, and see how this would affect studies of dark energy based on
baryonic oscillations.
We present the duration of the survey needed in order to constrain the equation of 
state of dark energy to the same accuracy.
Some of the results of this Section are summarised in Table.1.

If we relax the assumptions we have made regarding the evolution of the HI 
mass function, we could assume that 
the evolution
would be better described by Models A or B rather than C. 
In Fig.10 we plot the number density 
of HI sources for an integration time of
2 hours for Model A and an integration time of 
of 32 hours for Model B. We have chosen these integration times so that the number density 
of objects is enough to probe the Universe 
out to same redshift ($z_{\rm max}\sim1.5$, see Sec. 5.2)
as for Model C with tiled surveys with 4-hour integrations. 
We conclude that if we have made an error in choosing the 
evolution of the HI mass 
function the timescale of our survey would have to be multiplied by a factor in between 
$\sim 8$ and $\sim 0.5$ in 
order for us to get similar constraints. As discussed in Sec.4, 
Models A and
B are not preferred by observations, but that they are not yet 
completely ruled out:
it is plausible that a survey
would take between half of the time estimated and $\sim 8$-times longer.

\begin{figure}
\begin{center}
\centerline{\psfig{file=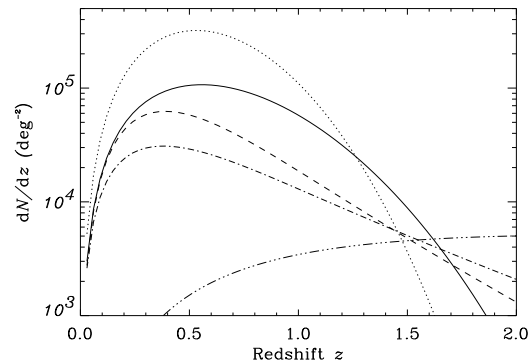,width=7.5cm,angle=0,clip=}}
\caption{The differential number density ${\rm d}N/{\rm d}z$
per ${\rm deg}^2$ of objects in an SKA survey with a signal-to-noise 
detection level of 10. Different integration times were chosen for our 
three evolution models: Model A (2-hour; dashed), B (32-hour; dotted) and C (4-hour; solid) 
and for a no-evolution model (6-hour;dot-dashed), probing out to same redshift
$z_{\rm max}\sim 1.5$. The 
triple-dot-dashed line defines the number density of objects needed at high redshift
for us to be cosmic variance limited when reconstructing the galaxy power spectrum (see Sec.5).}
\label{mass_limit}
\end{center}
\end{figure}

As we can see clearly from Sec.5 the gain which accrues from an increase in $FOV$ 
is linear in time. A survey with a telescope with 
twice the $FOV$ will be able to produce similar results 
in half of the time. The same is applicable to an increase in 
the effective bandwidth $\beta$ of the correlator. A telescope 
with $\beta$ halved will take twice as long
to produce a given survey. However, most of the volume surveyed 
is at high redshift, and given that the frequency range at low redshifts (0 to 0.5) 
is relatively large it may be desirable to neglect the low-redshift range of the survey
in order to produce a high-redshift survey with a higher value of $\beta$.

The relationship between the scaling of the field of view with frequency 
and performance of the survey is more 
complicated to assess as it changes depending on 
the $\beta \, FOV$ of the telescope but, on average, if a telescope is built with a 
field of view scaling with frequency as $\nu^{-1}$ 
it would be able to probe the same volume 
around 2-times slower than a telescope whose 
field of view scales as $\nu^{-2}$ (see Fig.9). 
If a telescope has a constant field of view it will be $\sim 4$-times 
slower than a telescope with a field of view scaling as $\nu^{-2}$ (see Fig.9).

The way in which HI galaxies trace the dark-matter fluctuations, the
bias, is another important uncertainty in our calculations. There are 
ways in which bias might make the baryonic oscillations method more
powerful than we have suggested. First, following
Blake \& Glazebrook (2004) a high bias population requires a lower number density of sources
to avoid shot-noise limitation. Second, as discussed in Sec 3.2, our method of 
estimating the HI mass function at high redshift effectively ignores the
most massive (i.e.\ elliptical) galaxies since they have low
HI content in the low-redshift Universe. The most massive systems at 
some (as yet unknown) high redshift are likely to have significant 
HI content and they are likely to be highly biased tracers of the dark-matter
mass as is seen for the most massive galaxies at low redshift (Norberg et al.
2001) and for high-redshift quasars (Croom et al.\ 2002). 

Another unknown in this analysis is the biasing model for the power spectrum. It is not known whether
a halo model (e.g. see Seljak 2000; Peacock \& Smith 2000) could induce a bias that could possible dilute some of 
the wiggles.
Blake \& Glazebrook (2003) argue that at large scales, where the first few baryonic wiggles are found, 
the power spectrum should have a constant bias as only extremely rare fluctuations will
be going non linear.

\section {Concluding remarks}

The most important features for the design of future radio telescopes have been identified
if they are to be used to probe dark energy in new and interesting ways.
The instantaneous 1.4-GHz {\em field of view} ($FOV$) of the telescope must be at 
least an order of magnitude
larger than the $\sim 1 ~ \rm deg^{2}$ fields 
of view achievable now by optical multi-object spectrograph
so that a `whole hemisphere' can be surveyed on a reasonable 
(i.e. $\sim 1 \, \rm yr$) 
timescale. HI surveys with the SKA would then in $\sim$ 1 yr contain
$\sim10^9$ galaxies with redshifts in the range 
$0 \ltsimeq z \ltsimeq 1.5$. 
The {\em sensitivity} of the telescope must eventually equal that proposed for the SKA, because 
this is the only way of detecting HI galaxies out to $z \sim 1.5$, and therefore obtaining
constraints on the dark-energy parameter $w$ of the order $\Delta w \sim 0.01$. An
SKA with a $FOV \sim 1 ~ \rm deg^{2}$ would be a very inefficient telescope for 
dark energy studies. The instantaneous {\em bandwidth} of the telescope
must cover the frequency range corresponding to 
HI in the redshift range $0.5 \ltsimeq z \ltsimeq 1.5$
in as few settings as possible, i.e.\ $\beta$, the ratio of the SKA bandwidth to the 
survey bandwidth, must be close to unity, because otherwise it trades off
linearly with $FOV$. The {\em baseline distribution} of the SKA must have a large
fraction (at least 50 per cent) of the collecting area within a $\sim 5$-km 
diameter core.

The most efficient way of making an SKA survey aimed at constraining the 
properties of dark energy
will be to first make a survey of all the sky available, and only then probe 
deeper in redshift. We have shown that, provided $\beta \, FOV \gtsimeq 10 \, \rm deg^2$,
HI surveys with a full SKA would, after $\sim \, 1 \rm \, yr$,
contain $\sim10^9 ~ (f_{\rm sky}/0.5)$ galaxies with redshifts in the redshift range $0 \ltsimeq z \ltsimeq 1.5$,
and hence provide constraints on the dark-energy parameter $w$ of order 
$\Delta w \simeq 0.01 (f_{\rm sky}/0.5)^{-0.5}$, 
where $f_{\rm sky}$ is the fraction of the whole sky.

\section {Acknowledgements}

We are very grateful for useful discussions with a large number of colleagues, most notably: 
Richard Battye, Chris Blake, Sarah Bridle, Greg Bryan, Chris Carilli, Peter Dewdney, Joe Silk, James Taylor, 
Thijs van der Hulst and Christian Wolf. 
We also thank the anonymous referee for numerous comments that have improved this paper. We thank the Gemini Project and PPARC
for a Studentship (FBA) and PPARC for a Senior Research Fellowship (SR).

\appendix

\section{Fitting formulae}

We present here (Table A1) some fitting formulae for the differential 
number density of objects expected in an SKA survey given the range of integration times 
shown in Table A1.

We have used the following fitting formula:

\begin{equation}
\frac{{\rm d}N}{{\rm d}z}={\rm A} \, z \,exp\left(-\frac{(z-z_c)^2}{2\Delta z^2}\right).
\end{equation}

\begin{table}
\begin{center}
\begin{tabular}{|l|c|c|c|}
\hline
Integration time & A & $z_c$ & $\Delta z$\\
\hline
1h  & $1.58 \, 10^5$ & 0.170 & 0.351 \\
4h & $2.52 \, 10^5$ & 0.211 & 0.461 \\
36h & $4.95 \, 10^5$ & 0.283 & 0.701 \\
360h & $9.33 \, 10^5$ & 0.386 & 1.045 \\
\end{tabular}
\end{center}
\caption{Fitting functions (dotted lines) for the differential number density 
of galaxies for a signal-to-noise detection of 10, a field of 
view scaling as $\nu^{-2}$ and for SKA-like sensitivity. 
We have chosen Model C to make our predictions (solid lines) and fitted for several integration times (see Table A1).} 
\end{table}

\begin{figure}
\begin{center}
\centerline{\psfig{file=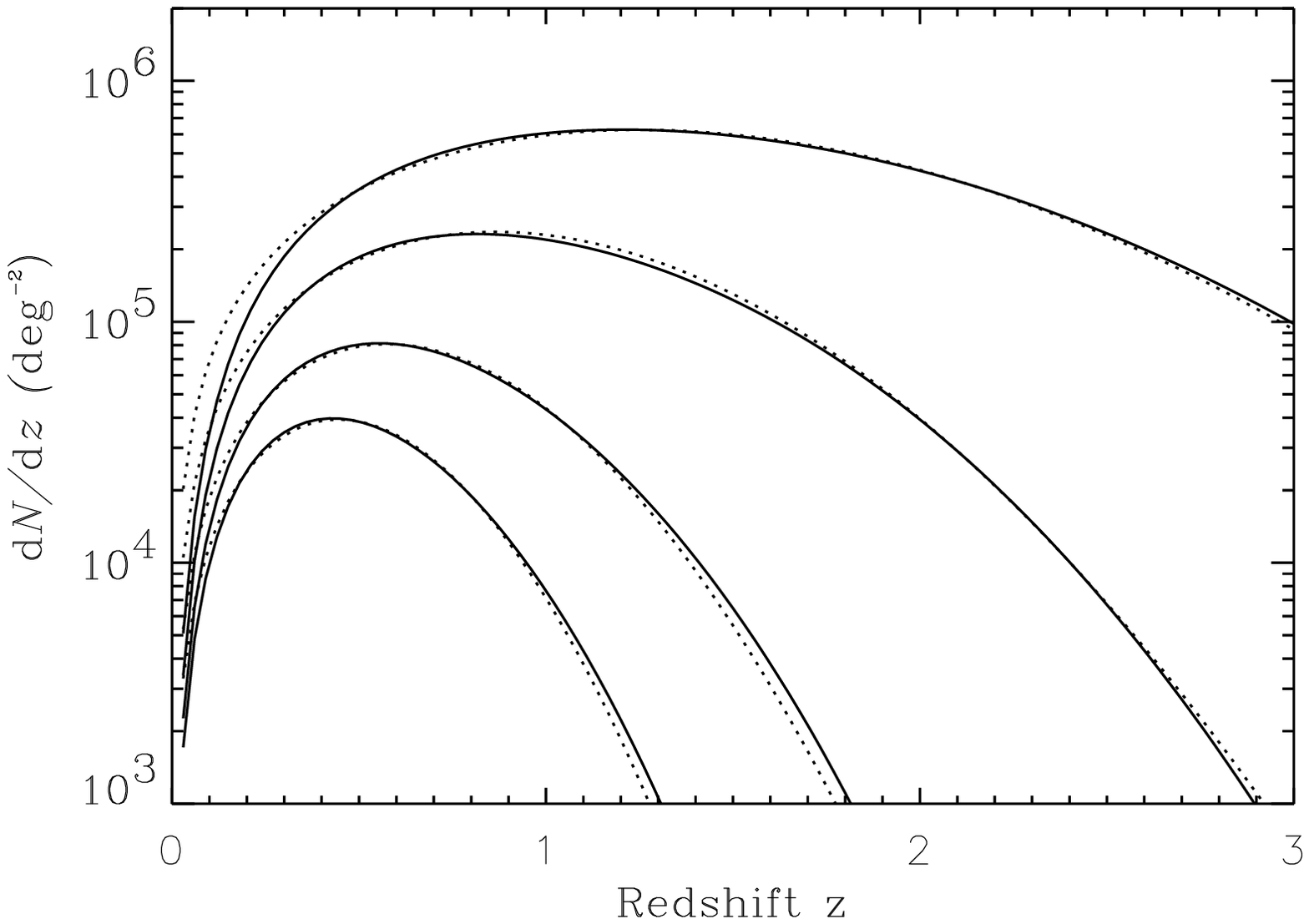,width=7.5cm,angle=0,clip=}}
\caption{
The comparison between our fitting formulae and the expected number density of sources
from Model C. For
all the plots we have assumed SKA-like sensitivity, 
a field of view scaling with frequency $\nu$ as $\nu^{-2}$, and a 
signal-to-noise detection limit of 10. The fittings 
are valid for the regions shown and become less accurate for redshifts larger than 3.0.  
}
\label{mass_limit}
\end{center}
\end{figure}

\end{document}